\documentclass[11pt,twoside]{article}
\usepackage[latin1]{inputenc}
\usepackage{times}
\usepackage{epsfig}
\usepackage{dsfont}
\usepackage{amsfonts}
\usepackage{float}
\usepackage{amsmath}
\usepackage{amsthm}
\usepackage{latexsym}
\usepackage{xr} 
\usepackage{natbib}
\usepackage{xcolor}
\usepackage{multirow}
\usepackage{rotating}
\usepackage{enumitem}

\relax 
\allowdisplaybreaks[3]

\newcommand{\mc}{\multicolumn}

\newcommand{\bay}{\begin{array}}
\newcommand{\eay}{\end{array}}

\newcommand{\bqa}{\begin{eqnarray*}}
\newcommand{\eqa}{\end{eqnarray*}}

\newcommand{\bee}{\begin{eqnarray*}}
\newcommand{\eee}{\end{eqnarray*}}

\newcommand{\bea}{\begin{eqnarray*}}
\newcommand{\eea}{\end{eqnarray*}}

\newcommand{\bqan}{\begin{eqnarray}}
\newcommand{\eqan}{\end{eqnarray}}

\newcommand{\be}{\begin{eqnarray}}
\newcommand{\ee}{\end{eqnarray}}

\newcommand{\bit}{\begin{itemize}}
\newcommand{\eit}{\end{itemize}}

\newcommand{\ben}{\begin{enumerate}}
\newcommand{\een}{\end{enumerate}}

\newcommand{\beq}{\begin{equation}}
\newcommand{\eeq}{\end{equation}}

\newcommand{\bdes}{\begin{description}}
\newcommand{\edes}{\end{description}}

\newcommand{\btb}{\begin{tabular}}
\newcommand{\etb}{\end{tabular}}

\newcommand{\bcen}{\begin{center}}
\newcommand{\ecen}{\end{center}}

\newcommand{\bmp}{\begin{minipage}}
\newcommand{\emp}{\end{minipage}}

\newcommand{\hsigma}{{\widehat \sigma}}

\newcommand{\vVh}{\widehat{\boldsymbol{V}}}

\newcommand{\vSigmah}{ \widehat{\vSigma}}

\newcommand{\vA}{\boldsymbol{A}}
\newcommand{\vB}{\boldsymbol{B}}

\newcommand{\vH}{\boldsymbol{H}}
\newcommand{\vI}{\boldsymbol{I}}
\newcommand{\vJ}{\boldsymbol{J}}

\newcommand{\vP}{\boldsymbol{P}}

\newcommand{\vT}{\boldsymbol{T}}

\newcommand{\vV}{\boldsymbol{V}}

\newcommand{\vX}{\boldsymbol{X}}

\newcommand{\valpha}{\boldsymbol{\alpha}}
\newcommand{\vbeta}{\boldsymbol{\beta}}
\newcommand{\vgamma}{\boldsymbol{\gamma}}

\newcommand{\vep}{\boldsymbol{\epsilon}}

\newcommand{\vmu}{\boldsymbol{\mu}}
\newcommand{\vnu}{\boldsymbol{\nu}}

\newcommand{\vSigma}{\boldsymbol{\Sigma}}

\newcommand{\veins}{{\bf 1}}
\newcommand{\vnull}{{\bf 0}}

\newcommand{\volX}{\boldsymbol{\overline{X}}}

\newcommand{\vwhD}{\boldsymbol{\widehat{D}}}

\newcommand{\vwhV}{\boldsymbol{\widehat{V}}}

\newcommand{\vwtJ}{\boldsymbol{\widetilde{J}}}

\newcommand{\vwtP}{\boldsymbol{\widetilde{P}}}

\newcommand{\rnc}{\renewcommand}
\newcommand{\nc}{\newcommand}
\newcommand{\mrm}{\mathrm}
\renewcommand{\b}{\textbf}

\nc{\mb}{\mathbb}
\nc{\mac}{\mathcal}
\nc{\E}{\mb{E}}
\nc{\N}{\mb{N}}
\nc{\R}{\mb{R}}
\nc{\Q}{\mb{Q}}
\rnc{\P}{\mrm P}
\nc{\bP}{\b P}
\nc{\bA}{\b A}
\rnc{\d}{\mrm d}
\nc{\C}{\mc{C}}
\nc{\D}{\mc{D}}
\nc{\B}{\mc{B}}
\nc{\I}{\mc I}
\nc{\J}{\mc J}
\nc{\JI}{{\J\I}}
\rnc{\IJ}{{\I\J}}
\nc{\ji}{{\J|\I}}
\nc{\gDg}{\stackrel{d}{=}}
\nc{\oPo}{\stackrel{\mrm p}{\longrightarrow}}
\nc{\oWo}{\stackrel{w}{\longrightarrow}}
\nc{\oDo}{\stackrel{d}{\longrightarrow}}
\nc{\nae}{Nelson-Aalen estimator}
\nc{\aje}{Aalen-Johansen estimator}
\nc{\naeL}{Nelson-Aalen estimator\ }
\nc{\ajeL}{Aalen-Johansen estimator\ }
\nc{\CIF}{cumulative incidence function}
\nc{\CIFL}{cumulative incidence function\ }
\nc{\wh}{\widehat}

\newcommand{\bks}{\bigoplus}

\newcommand{\Cov}{\operatorname{{\it Cov}}}
\newcommand{\Var}{\operatorname{{\it Var}}}

\definecolor{darkolivegreen}{rgb}{0.33, 0.42, 0.18}
\definecolor{darkgoldenrod}{rgb}{0.72, 0.53, 0.04}

\usepackage{wrapfig,lipsum,booktabs,makecell}
\usepackage{fancyvrb}

\let\proglang=\textsc
\newcommand{\pkg}[1]{{\fontseries{b}\selectfont #1}}
\let\code=\texttt

\DefineVerbatimEnvironment{Code}{Verbatim}{}
\DefineVerbatimEnvironment{CodeInput}{Verbatim}{fontshape=sl}
\DefineVerbatimEnvironment{CodeOutput}{Verbatim}{}

\advance\textheight by \topskip
\begin{document}


\title{\Large \bf Analysis of Multivariate Data and Repeated Measures Designs with the \textsc{R} Package \pkg{MANOVA.RM}}

\author{Sarah Friedrich$^{*}$, Frank Konietschke$^{**}$ and  Markus Pauly$^{*}$ \\[1ex] 
}
\maketitle

\begin{abstract}
The numerical availability of statistical inference methods for a modern and robust analysis of longitudinal- and multivariate data in factorial experiments is 
an essential element in research and education. While existing approaches that rely on specific distributional assumptions of the data (multivariate normality and/or 
characteristic covariance matrices) are implemented in statistical software packages, there is a need for user-friendly software that can be used for the analysis of data that 
do not fulfill the aforementioned assumptions and provide accurate $p$-value and confidence interval estimates. Therefore, newly developed statistical
methods for the analysis of repeated measures designs and multivariate data that neither assume multivariate normality nor specific covariance matrices have  been implemented in the freely available \textsc{R}-package \pkg{MANOVA.RM}. The package is equipped with a graphical user interface for plausible applications in academia and other educational purpose. Several motivating examples
illustrate the application of the methods. 

\end{abstract}


\vfill
\vfill

\noindent${}^{*}$ {Ulm University, Institute of Statistics, Germany\\
 \mbox{ }\hspace{1 ex}email: sarah.friedrich@uni-ulm.de}\\

\noindent${}^{**}$ {University of Texas at Dallas, USA}\\

\newpage

\section{Introduction}

Nowadays, a large amount of measurements are taken per experimental unit or subject in many experimental studies\textemdash requiring inferential methods from multivariate analysis in a unified way. 
Here we distinguish between {\it two cases}: 
\begin{enumerate}
\item If the same quantity is measured under different treatment conditions or at different time points, a {\it repeated measures (RM)} design is present. 
Therein,  observations are measured on the same scale and are {\it combinable}.
\item  If different quantities are measured on the same unit or subject, a {\it multivariate analysis of variance (MANOVA)} design  is apparent. In such a situation, data is measured on different scales and not combinable (e.g. height and weight). 
\end{enumerate}
These two different definitions do not only lead to different questions of interest but also require different inference procedures as outlined below. 
{\color{black} In particular, the main difference between the two approaches is that in repeated measures designs comparisons between the response variables are meaningful. This means that also hypotheses regarding sub-plot factors (e.g. time) are of interest. On the other hand, MANOVA settings are usually only designed to detect effects of the observed factors (and interactions thereof) on the multivariate outcome vectors.} 

Despite their differences, MANOVA- and RM-type techniques share the same advantages over classical univariate endpoint-wise\textemdash ANOVA-type\textemdash analyses: 
\bit 
\item They provide  joint inference and take the  dependency across the endpoints into account, thus leading to possibly larger power to detect underlying effects. 
\item They allow for testing of additional factorial structures and 
\item can easily be equipped with a closed testing procedure for subsequently detecting local effects in specific components. 
\eit

Focusing on metric data and mean-based procedures, MANOVA and RM-models are typically inferred by means of 
``classical'' procedures such as Wilks' Lambda, Lawley-Hotelling, Roy's largest root \citep{Davis,johnson,anderson2001new} or (generalized) linear mixed models with generalized estimating equations. For the classical one-way layout, these methods are implemented in \textsc{R} within the \code{manova} function in the \pkg{stats} package, where one can choose between the options `Pillai`,  `Wilks`, `Hotelling-Lawley` and `Roy`.
Nonparametric rank-based methods for null hypotheses formulated in distribution functions are implemented within the packages \pkg{npmv} for one- and two-way MANOVA \citep{npmv} and \pkg{nparLD} for several repeated measures designs \citep{nparLD}. In case of fixed block effects, the \pkg{GFD} package \citep{friedrichGFD}, which implements a permutation Wald-type test in the univariate setting, can also be used.

(Generalized) linear mixed models are implemented in the \code{lm} and the \code{glm} function \citep[package \pkg{stats},][]{R} for univariate data as well as in the \pkg{SCGLR} package for Generalized Linear Model estimation in the context of multivariate data \citep{SCGLR}. The \code{Anova} and \code{Manova} function in the \pkg{car} package \citep{Fox} calculate type-II and type-III analysis-of-variance tables for objects produced by, e.g., \code{lm}, \code{glm} or \code{manova} in the univariate and multivariate context, respectively. In the MANOVA context, repeated measures designs can be included as well.

Most of these procedures, however, rely on specific distributional assumptions (such as multivariate normality) and/or specific covariance or correlation structures (e.g., homogeneity between groups or, for RM,  compound symmetry; possibly implying equal correlation between measurements) which may often not be justifiable in real data. In particular, with decreasing sample sizes and increasing dimensions, such presumptions are almost impossible to verify in practice and may lead to inflated type-I-errors, cf. 
\cite{vallejo2001effects, lix2004multivariate, vallejo2007comparative, livacic2010analysis}. 
To this end, several alternative procedures have been developed that tackle the above problems and have been compared in extensive simulation studies, see amongst others \citet{brunner:2001, lix2007comparison, gupta2008manova, zhang2011two, harrar2012modified, Kon:2015, xiao2016modified, bathke2016using, mcfarquhar2016multivariate, friedrich2017permuting, livacic2017power, friedrichMATS} and the references cited therein.
Here, we focus on statistical methods that are valid in the multivariate Behrens-Fisher situation\textemdash equal covariance matrices across the groups is not assumed\textemdash and provide accurate inferential results in terms of p-value estimates and confidence intervals for the parameters of interest. In particular, general Wald-type test statistics (for MANOVA and RM), ANOVA-type statistics (for RM) and modified ANOVA-type tests (for MANOVA) are implemented in \pkg{MANOVA.RM} because they
\bit
\item can be used to test hypotheses in various factorial designs in a flexible way, 
\item  their sampling distribution can be approximated by resampling techniques, even allowing their application for small sample sizes,
\item and are appropriate methods in the Behrens-Fisher situation. 
\eit

To make the methods freely accessible we have provided the \textsc{R} package \pkg{MANOVA.RM} for daily statistical analyses. It is available from the \textsc{R} Archive at 
\begin{center}
{\it https://CRAN.R-project.org/package=MANOVA.RM}
\end{center}
The main functions \code{RM} (for RM designs) and \code{MANOVA} (for MANOVA designs) are developed in style of the well known ANOVA functions \code{lm}
or \code{aov} \citep[\textsc{R} package \pkg{stats},][]{R}. Its user-friendly application not only provides the $p$-values and test statistics of interest but also 
a descriptive overview together with component-wise two-sided confidence intervals. Moreover, the \code{MANOVA} function even allows for an easy 
calculation and confidence ellipsoids plots for specified multivariate contrasts as described in \citet{friedrichMATS}. 

Specifically, for testing multivariate main- and interaction effects in one-, two- and higher-way MANOVA models, the \code{MANOVA} function provides the 
\bit
\item Wald-type statistic (WTS) proposed by \citet{Kon:2015} using a parametric bootstrap, a wild bootstrap or its asymptotic $\chi^2$-distribution for $p$-value computations, and
\item the modified ANOVA-type statistic (MATS) proposed by \citet{friedrichMATS} using a parametric or wild bootstrap procedure for $p$-value computations.
\eit
In addition to multivariate group-wise effects, the \code{RM} function also allows to test hypotheses formulated across sub-plot factors. The implemented test statistics are
\bit
\item the ANOVA-type statistic (ATS) using an $F$-approximation as considered in \citet{brunner:2001} as well as a parametric and a wild bootstrap approach and
\item the Wald-type statistic (WTS) using the asymptotic $\chi^2$-distribution \citep{brunner:2001}, the permutation technique proposed in \citet{friedrich2017permuting} as well as a parametric \citep{bathke2016using} and a wild bootstrap approach for $p$-value estimation.
\eit

Th{e} paper is organized as follows: In Section~\ref{Model} the multivariate statistical model as well as the implemented inference procedures are described. 
The application of the \textsc{R} package \pkg{MANOVA.RM} is exemplified on several Repeated Measures and MANOVA Examples in Section~\ref{Examples}. 
Finally, the paper closes with a discussion in Section~\ref{Conclusion}.

Throughout the paper we use the subsequent notation from multivariate linear models: For $a\in \N$ we denote by $\vP_a = \vI_a - \frac{1}{a} \vJ_a$ the $a$-dimensional centering matrix, 
by $\vI_a$ the $a$-dimensional identity matrix and by $\vJ_a$ the $a \times a$ matrix of 1's, i.e., $\vJ_a = \boldsymbol{1}_a \boldsymbol{1}_a'$, where $\boldsymbol{1}_a=(1, \dots, 1)'$ is the $a$-dimensional column vector of 1's.

\section{Statistical model and inference methods}\label{Model}

For both the RM and the MANOVA design being equipped with an arbitrary number of fixed factors, we consider the general linear model given by $d$-variate random vectors
\bqan\label{model}
\vX_{ik} \ = \ (X_{ijk})_{j=1}^{d} &= \vmu_i + \vep_{ik}.
\eqan
Here, $k=1, \dots,n_i$ denotes the experimental unit or subject in group $i=1, \dots, a$. 
Note, that a higher-way factorial structure on the groups/whole-plots or sub-plots can be achieved by sub-indexing the indices $i$ (group/whole-plot) or $j$ (sub-plot) into $i_1,\dots,i_p$ or $j_1,\dots,j_q$.
In this model $\vmu_i = (\mu_{i1}, \dots, \mu_{id})' \in \mathbb{R}^d$ is the mean vector in group $i=1, \dots, a$ and for each fixed $i$ it is assumed that the error terms $\vep_{ik}, k=1,\dots,n_i,$ are independent and identically distributed $d$-variate random vectors with mean $E(\vep_{i1}) = 0$ and existing 
variances $0 < \sigma_{ij}^2 = \Var(X_{ijk}) < \infty , ~ j=1, \dots, d$. For the WTS-type procedures we additionally assume positive definite covariance matrices 
$\Cov(\vep_{i1})=\vV_i > 0$ and existing finite fourth moments $E(||\vep_{i1}||^4) < \infty.$\\

In this model, hypotheses for RM or MANOVA can be formulated by means of an adequate contrast hypothesis matrix $\vH$ by
$$
 H_0: \vH \vmu = \vnull.
$$

Let $\volX_{\bullet} = (\volX_{1\cdot}', \dots, \volX_{a\cdot}')'$ denote the vector of pooled group means $\volX_{i\cdot} = \frac{1}{n_i}\sum_{k=1}^{n_i}\vX_{ik}, i=1, \dots, a$ and  $\vSigmah_N = \oplus_{i=1}^a N \vVh_i/n_i$ the estimated covariance of $\sqrt{N} \volX_{\bullet}$. Here, $N=\sum_i n_i$ and $\vVh_i = \frac{1}{n_i-1}\sum_{k=1}^{n_i}(\vX_{ik} - \volX_{i \cdot})(\vX_{ik} - \volX_{i \cdot})'$. In this set-up \citet{Kon:2015} propose 

a Wald-type statistic (WTS) 
\bqan \label{WTS}
T_N = T_N(\vX) =   N \volX_{\bullet}' \vT (\vT \vSigmah_N \vT)^+\vT \volX_{\bullet},
\eqan
for testing $H_0$, where $\vT = \vH'(\vH \vH')^+\vH$, $\vX=\{\vX_{11},\dots,\vX_{an_a}\}$, and $\vA^+$ denotes the Moore-Penrose inverse of the matrix $\vA$. Since its asymptotic $\chi_{rank(\vT)}^2$ null distribution provides a poor finite sample approximation, they propose the following asymptotic model-based bootstrap approach: Given the data $\vX$ let $\vX_{ik}^\star \sim N(\vnull, \vVh_i), i=1,\dots, a, k=1,\dots,n_i,$ be independent random vectors that are used for recalculating the test statistic as $T_N^\star = T_N(\vX^\star)$, where $\vX^\star=\{\vX_{11}^\star,\dots,\vX_{an_a}^\star\}$. Denoting by $c^\star$ the corresponding $(1-\alpha)$-quantile of the (conditional) distribution of $T_N^\star$ the test rejects $H_0$ if $T_N > c^\star$. The validity of this procedure (also named parametric bootstrap WTS) is proven in \citet{Kon:2015}. 

This procedure is not only applicable for MANOVA but also for RM designs (\citet{bathke2016using}). However, \citet{friedrich2017permuting} proposed a more favourable technique for RM designs. 
It is based on an at first blush chaotic resampling method: Wild permutation of all pooled components without taking group membership or possible dependencies into account. Denoting the resulting permuted data set as $\vX^\pi$ their permutation test for RM models rejects $H_0$ if $T_N > c^\pi$. Here $c^\pi$ denotes the $(1-\alpha)$-quantile of the (conditional) distribution of the permutation version of the test statistic $T_N^\pi = T_N(\vX^\pi)$. As shown in extensive simulations in \citet{friedrich2017permuting} and the corresponding supporting information this 'wild' permutation WTS method controls the type-1 error rate very well. Note that this procedure is only applicable for RM due to the commensurate nature of their components. In MANOVA set-ups the permutation would stir different scalings making comparisons meaningless. 

In addition to these WTS procedures two other statistics are considered as well. For RM the well-established ANOVA-type statistic (ATS) 
\bqan \label{ATS}
{Q}_N = N \volX_{\bullet}' \vT \volX_{\bullet}
\eqan
by \citet{brunner:2001} is implemented together with the enhanced $F$-approximation of the statistic proposed by \citet{BrDeMu:1997,BrunnerDomhofLanger2002} and also implemented in the SAS PROC Mixed procedure. Although known to be rather conservative it has the advantage (over the WTS) of being applicable in case of eventually singular covariance matrices $\vV_i$ or $\vwhV_i$ since it waives the Moore-Penrose inverse involved in \eqref{WTS}. 

Similar to the permuted WTS the ATS given in \eqref{ATS} is only applicable for RM since it is not invariant under scale transformations (e.g. change of units) of the univariate components. To nevertheless provide a robust method for MANOVA settings which is also applicable in case of singular $\vV_i$ or $\vwhV_i$, \citet{friedrichMATS} have recently proposed the novel MATS (modified ATS)
\bqan \label{modATS}
M_N = M_N(\vX) = N \volX_{\bullet}' \vT (\vT \vwhD_N \vT)^+\vT \volX_{\bullet}.
\eqan
Here, the involved diagonal matrix $\vwhD_N = \oplus_{1\leq i\leq a, 1\leq s\leq d}{N}\hsigma_{is}^2/n_i$ of the empirical variances $\hsigma_{is}^2$ of component $s$ in group $i$, deduces an invariance under component-wise scale transformations of the MATS. To obtain an accurate finite sample performance, it is also equipped with an asymptotic model based bootstrap approach. That is, MATS rejects $H_0$ if $M_N > \tilde{c}^\star$, where $\tilde{c}^\star$ is the $(1-\alpha)$-quantile of the (conditional) distribution of the bootstrapped statistic $M_N^\star = M_N(\vX^\star)$. 
In addition, we implemented a wild bootstrap approach, which is based on multiplying the centered data vectors $(\vX_{ik} - \vX_{i \cdot})$ with random weights $W_{ik}$ fulfilling $E(W_{ik})=0, \Var(W_{ik})=1$ and $\sup_{i, k}E(W_{ik}^4)<\infty$. In the package, we implemented Rademacher distributed weights, i.e., random signs.
Extensive simulations in \citet{friedrichMATS} not only confirm its applicability in case of singular covariance matrices but also disclose a very robust behaviour that even seems to be advantageous over the parametrically bootstrapped WTS of \citet{Kon:2015}. However, both procedures, as well as the 'usual' asymptotic WTS are displayed within the MANOVA functions. 
{All of the aforementioned procedures are applicable in various factorial designs in a unified way, i.e. when more than one factor may impact the response. The specific models and the hypotheses being tested will be discussed in the next section.


\subsection{Special Designs and Hypotheses}
In order to provide a general overview of different statistical designs and layouts that can be analyzed with \pkg{MANOVA.RM} we exemplify few designs that occur frequently in practical applications and discuss 
the model, hypotheses and limitations. All of the methods being implemented in \pkg{MANOVA.RM} are even applicable in higher-way layouts than being presented here; and the list should not be seen as the limited application of the package. The models are derived by sub-indexing the index $i$ in model (\ref{model}) $\vX_{ik} = \vmu_i + \vep_{ik}$ in the following ways:

\begin{itemize}
  \item{\bf One-Way ($\vA$):} Writing $\vmu_i = \vnu+\valpha_{i}$ we have $\vX_{ik}=\vnu+\valpha_{i}+\vep_{ik}$ with $\sum_{i=1}^a \valpha_i=\vnull$ and obtain the null hypothesis of 'no group' or 'factor $\vA$' effect as
  \bqa
  H_0(\vA):\{(\vP_a\otimes \vI_d) \vmu = \vnull\} &=& \{\vmu_1=\dots=\vmu_a\}\\ &=& \{\valpha_1=\dots=\valpha_a=\vnull\}.
  \eqa
  In case of $a=2$ this includes the famous {\it multivariate Behrens-Fisher problem} as, e.g., analyzed in \citet{yao1965approximate,nel1986solution,christensen1997comparison,krishnamoorthy2004modified} or \citet{yanagihara2005three}.
  \item {\bf Crossed Two-Way ($\vA\times \vB$):}
  Splitting the index into two  and writing $\vmu_{ij} = \vnu+\valpha_i+\vbeta_j+\vgamma_{ij}$ we obtain the model  $\vX_{ijk}= \vnu+\valpha_i+\vbeta_j+\vgamma_{ij}+\vep_{ijk}, \,  1\leq i\leq a,~1\leq j\leq b,~ 1\leq k\leq n_{ij}$
  with $\sum_i{\valpha_i}=\sum_j{\vbeta_j}=\sum_i{\vgamma_{ij}}=\sum_j{\vgamma_{ij}}=\boldsymbol{0}.$ The corresponding null hypotheses of no main effects in $\vA$ or $\vB$ and no interaction effect between $\vA$ and $\vB$ can be written as:
\bqa
  H_0(\vA) :\{(\vP_a\otimes{b}^{-1}~\vJ_b\otimes\vI_d)~\vmu = \vnull\}  &=& \{\valpha_1=\dots=\valpha_a=\vnull\},\\
  H_0(\vB) :\{(a^{-1}~\vJ_a\otimes\vP_b\otimes\vI_d)~\vmu = \vnull\}  &=& \{\vbeta_1=\dots=\vbeta_b=\vnull\},\\
  H_0(\vA\vB):\{(\vP_a\otimes\vP_b\otimes\vI_d)~\vmu=\vnull\}\;\;\;\;\;\;\,&=& \{\vgamma_{11}=\dots=\vgamma_{ab}=\vnull\}.
  \eqa
  \item {\bf Hierarchically nested Two-Way ($\vB(\vA)$):} A fixed subcategory $B$ within factor $A$ can be introduced via the model $\vX_{ijk}= {\vnu+\valpha_i+\vbeta_{j(i)}} +  \vep_{ijk}, 1\leq i\leq a,~1\leq j\leq b_i~,1\leq k\leq n_{ij}$
  with $\sum_i{\valpha_i}=\sum_j{\vbeta_{j(i)}}=\boldsymbol{0}.$ Here, the hypotheses of no main effect $A$ or no sub-category main effect $B$ can be written as
  \bqa
  H_0(A) :\{(\vP_a\tilde{\vJ}_b\otimes\vI_d)~\vmu = \vnull\}  &=& \{\valpha_1=\dots=\valpha_a=\vnull\},\\
  H_0(B(A)) :\{ (\vwtP_b\otimes \vI_d)~\vmu=  \vnull\}  &=& \{\vbeta_{j(i)} = 0 \forall\ 1\leq i\leq a, 1\leq j\leq b_i\}
  \eqa
with $\vwtP_b=\bks_{j=1}^a\vP_{b_j},~\vwtJ_b=\bks_{j=1}^a b_j^{-1}\veins_{b_j}'$ and $\vmu:=(\vmu_{11}',...,\vmu_{1b_1}',\vmu_{21}',...,\vmu_{2b_2}',...,\vmu_{ab_a}')'$.

We only implemented balanced designs, i.e., $b_i=b$ for all $i=1, \dots, a$. Hierarchically nested three-way designs or arbitrary crossed higher-way layouts can be introduced similarly and are implemented as well.
  \item {\bf Repeated Measures and Split Plot Designs} are covered by setting $d=t$, where even hypotheses about sub-plots can be formulated. We exemplify this for profile analyses in the special case of a one-sample RM design with $a=1$ 
  \bqa
  H_0(\text{Time}): \{\vP_t~\vmu=\boldsymbol{0}\} &=& \{\mu_{11}=\dots=\mu_{1t}\},
  \eqa
  as well as for a two-sample RM design with $a=2$:
  \bqa
 H_0(\text{Parallel}): \{\vT_P~\vmu=\boldsymbol{0}\} &=&\{\vmu_1-\vmu_2=\gamma~\veins_t ~~\text{for some}~~\gamma \in \mathbb{R}\}\\
H_0(\text{Flat}): \{\vT_F~\vmu=\boldsymbol{0}\}&=& \{\vmu_{1s}+\vmu_{2s}=\bar{\vmu}_{1\cdot}+\bar{\vmu}_{2\cdot} ~~\text{for all}~~s\} \\
H_0(\text{Identical}): \{\vT_I~\vmu=\boldsymbol{0}\} &=& \{\vmu_1=\vmu_2\}
\eqa\\[-1.5ex]
    with $\vT_F=\vP_t~(\vI_t~\vdots~\vI_t)$, $\vT_P=(\veins_{t-1}~\vdots-\vI_{t-1})~\vT_I$ and $\vT_I = (\vI_t~\vdots-\vI_t)$.\\[1ex]
\begin{center}
\includegraphics[scale=0.8]{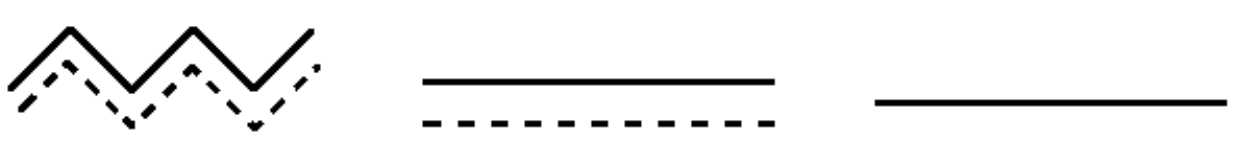}
\end{center}
\vspace{-1ex}{\hspace{2.9cm}parallel\hspace{2.7cm} flat \hspace{2.7cm} identical}\\[1ex]
Note, that we could also employ more complex factorial structures on the repeated measurements (i.e., more sub-plot factors) by splitting up the index $j$.

\end{itemize}

}

\section{Examples} \label{Examples}
 
To demonstrate the use of the \code{RM} and the \code{MANOVA} function, we provide several examples for both repeated measures designs and multivariate data in the following. 
Furthermore, {\color{black}the \pkg{MANOVA.RM} package} is equipped with an optional GUI, based on \pkg{RGtk2}~\citep{RGtk2}, which will be explained in detail in Section \ref{sec:GUI} below.

\subsection{Repeated Measures Designs}

The function \code{RM} returns {\color{black}an object of class \code{RM}} from which the user may obtain plots and summaries of the results using \code{plot()}, \code{print()} and \code{summary()}, respectively. 
Here, \code{print()} returns a short summary of the results, i.e., the values of the test statistics along with degrees of freedom and corresponding $p$~values whereas \code{summary()} also displays some descriptive statistics such as the means, sample sizes and confidence intervals for the different factor level combinations. Plotting is based on \pkg{plotrix}~\citep{plotrix}. {\color{black} For two- and higher-way layouts, the factors for plotting can be additionally specified in the \code{plot} call, see the examples below.}
\begin{CodeInput}
R> RM(formula, data, subject, no.subf = 1, iter = 10000, 
+      alpha = 0.05, resampling = "Perm", CPU, seed, 
+      CI.method = "t-quantile", dec = 3)
\end{CodeInput}

Data need to be provided in long format, i.e., one row per measurement.
Here, \code{subject} specifies the column name of the subjects variable in the data frame, while \code{no.subf} denotes the number of sub-plot factors considered. The number of cores used for parallel computing as well as a random seed can optionally be specified using \code{CPU} and \code{seed}, respectively. For calculating the confidence intervals, the user can choose between $t$-quantiles (the default) and the quantiles based on the resampled WTS. The results are rounded to \code{dec} digits. 

\subsubsection{Example 1: One whole-plot and two sub-plot factors}

For illustration purposes, we consider the data set \code{o2cons}, which is included in \pkg{MANOVA.RM}. This data set contains measurements of the oxygen consumption of leukocytes in the presence and absence of inactivated staphylococci at three consecutive time points. Due to the study design, both time and staphylococci are sub-plot factors while the treatment (Verum vs. Placebo) is a whole-plot factor \citep{friedrich2017permuting}.

\begin{CodeInput}
R> data("o2cons")
R> model1 <- RM(O2 ~ Group * Staphylococci * Time, data = o2cons, 
+	       subject = "Subject",  no.subf = 2, iter = 1000, 
+               resampling = "Perm", seed = 1234)
R> summary(model1)
\end{CodeInput}
\begin{CodeOutput}
	
 Call: 
 O2 ~ Group * Staphylococci * Time
 
Descriptive:
   Group Staphylococci Time  n Means  Lower 95 % CI  Upper 95 % CI
1      P             0    6 12 1.322         1.150         1.493
5      P             0   12 12 2.430         2.196         2.664
9      P             0   18 12 3.425         3.123         3.727
3      P             1    6 12 1.618         1.479         1.758
7      P             1   12 12 2.434         2.164         2.704
11     P             1   18 12 3.527         3.273         3.781
2      V             0    6 12 1.394         1.201         1.588
6      V             0   12 12 2.570         2.355         2.785
10     V             0   18 12 3.677         3.374         3.979
4      V             1    6 12 1.656         1.471         1.840
8      V             1   12 12 2.799         2.500         3.098
12     V             1   18 12 4.029         3.802         4.257

Wald-Type Statistic (WTS):
                         Test statistic df  p-value
Group                            11.167  1   0.001
Staphylococci                    20.401  1   0.000
Group:Staphylococci               2.554  1   0.110
Time                           4113.057  2   0.000
Group:Time                       24.105  2   0.000
Staphylococci:Time                4.334  2   0.115
Group:Staphylococci:Time          4.303  2   0.116

ANOVA-Type Statistic (ATS):
                          Test statistic df1    df2   p-value
Group                            11.167 1.000 316.278   0.001
Staphylococci                    20.401 1.000     Inf   0.000
Group:Staphylococci               2.554 1.000     Inf   0.110
Time                            960.208 1.524     Inf   0.000
Group:Time                        5.393 1.524     Inf   0.009
Staphylococci:Time                2.366 1.983     Inf   0.094
Group:Staphylococci:Time          2.147 1.983     Inf   0.117
 
p-values resampling:
                            Perm (WTS) Perm (ATS)
Group                         0.005         NA
Staphylococci                 0.000         NA
Group:Staphylococci           0.138         NA
Time                          0.000         NA
Group:Time                    0.000         NA
Staphylococci:Time            0.161         NA
Group:Staphylococci:Time      0.163         NA
\end{CodeOutput}

The output consists of four parts: \code{model1\$Descriptive} gives an overview of the descriptive statistics: The number of observations, mean and confidence intervals are displayed for each factor level combination. Second, \code{model1\$WTS} contains the results for the Wald-type test: The test statistic, degree of freedom and $p$-values based on the asymptotic $\chi^2$-distribution are displayed. Note that the $\chi^2$-approximation is very liberal for small sample sizes, cf. \citet{Kon:2015, friedrich2017permuting}. The corresponding results based on the ATS are contained within \code{model1\$ATS}. This test statistic tends to rather conservative decisions in case of small sample sizes and is even asymptotically only an approximation, thus not providing an asymptotic level $\alpha$ test, see  \citet{brunner:2001, friedrich2017permuting}. Finally, \code{model1\$resampling} contains the $p$-values based on the chosen resampling approach. For the ATS, the permutation approach is not feasible since it would result in an incorrect covariance structure, and is therefore not implemented. Due to the above mentioned issues for small sample sizes, the respective resampling procedure is recommended for such situations.

In this example, we find significant effects of all factors as well as a significant interaction between group and time.

\subsubsection{Example 2: Two sub-plot and two whole-plot factors}
We consider the data set \code{EEG} from the \pkg{MANOVA.RM} package: At the Department of Neurology, University Clinic of Salzburg, 160 patients were diagnosed with either AD, MCI, or SCC, based on neuropsychological diagnostics \citep{bathke2016using}. This data set contains $z$-scores for brain rate and Hjorth complexity, each measured at frontal, temporal and central electrode positions and averaged across hemispheres. In addition to standardization, complexity values were multiplied by $-1$ in order to make them more easily comparable to brain rate values: For brain rate we know that the values decrease with age and pathology, while Hjorth complexity values are known to increase with age and pathology. The three whole-plot factors considered were sex (men vs. women), diagnosis (AD vs. MCI vs. SCC), and age ($<70$ vs. $\geq70$ years). Additionally, the sub-plot factors region (frontal, temporal, central) and feature (brain rate, complexity) structure the response vector.

\begin{CodeInput}
R> data("EEG")
R> EEG_model <- RM(resp ~ sex * diagnosis * feature * region, 
+               data = EEG, subject = "id", no.subf = 2, 
+               resampling = "WildBS", iter = 1000,  alpha = 0.01,
+               CPU = 4, seed = 123)
R> summary(EEG_model)
\end{CodeInput}

\begin{CodeOutput}
Call: 
resp ~ sex * diagnosis * feature * region

Descriptive:
   sex diagnosis   feature   region   n  Means  Lower 99 % CI Upper 99 % CI
1    M        AD  brainrate  central 12 -1.010        -4.881         2.861
13   M        AD  brainrate  frontal 12 -1.007        -4.991         2.977
25   M        AD  brainrate temporal 12 -0.987        -4.493         2.519
7    M        AD complexity  central 12 -1.488       -10.053         7.077
19   M        AD complexity  frontal 12 -1.086        -6.906         4.735
31   M        AD complexity temporal 12 -1.320        -7.203         4.562
3    M       MCI  brainrate  central 27 -0.447        -1.591         0.696
15   M       MCI  brainrate  frontal 27 -0.464        -1.646         0.719
27   M       MCI  brainrate temporal 27 -0.506        -1.584         0.572
9    M       MCI complexity  central 27 -0.257        -1.139         0.625
21   M       MCI complexity  frontal 27 -0.459        -1.997         1.079
33   M       MCI complexity temporal 27 -0.490        -1.796         0.816
5    M       SCC  brainrate  central 20  0.459        -0.414         1.332
17   M       SCC  brainrate  frontal 20  0.243        -0.670         1.156
29   M       SCC  brainrate temporal 20  0.409        -1.210         2.028
11   M       SCC complexity  central 20  0.349        -0.070         0.767
23   M       SCC complexity  frontal 20  0.095        -1.037         1.227
35   M       SCC complexity temporal 20  0.314        -0.598         1.226
2    W        AD  brainrate  central 24 -0.294        -1.978         1.391
14   W        AD  brainrate  frontal 24 -0.159        -1.813         1.495
26   W        AD  brainrate temporal 24 -0.285        -1.776         1.206
8    W        AD complexity  central 24 -0.128        -1.372         1.116
20   W        AD complexity  frontal 24  0.026        -1.212         1.264
32   W        AD complexity temporal 24 -0.194        -1.670         1.283
4    W       MCI  brainrate  central 30 -0.106        -1.076         0.863
16   W       MCI  brainrate  frontal 30 -0.074        -1.032         0.885
28   W       MCI  brainrate temporal 30 -0.069        -1.064         0.925
10   W       MCI complexity  central 30  0.094        -0.464         0.652
22   W       MCI complexity  frontal 30  0.131        -0.768         1.031
34   W       MCI complexity temporal 30  0.121        -0.652         0.895
6    W       SCC  brainrate  central 47  0.537        -0.049         1.124
18   W       SCC  brainrate  frontal 47  0.548        -0.062         1.159
30   W       SCC  brainrate temporal 47  0.559        -0.015         1.133
12   W       SCC complexity  central 47  0.384         0.110         0.659
24   W       SCC complexity  frontal 47  0.403        -0.038         0.845
36   W       SCC complexity temporal 47  0.506         0.132         0.880

Wald-Type Statistic (WTS):
                             Test statistic df  p-value
sex                                   9.973  1   0.002
diagnosis                            42.383  2   0.000
sex:diagnosis                         3.777  2   0.151
feature                               0.086  1   0.769
sex:feature                           2.167  1   0.141
diagnosis:feature                     5.317  2   0.070
sex:diagnosis:feature                 1.735  2   0.420
region                                0.070  2   0.966
sex:region                            0.876  2   0.645
diagnosis:region                      6.121  4   0.190
sex:diagnosis:region                  1.532  4   0.821
feature:region                        0.652  2   0.722
sex:feature:region                    0.423  2   0.810
diagnosis:feature:region              7.142  4   0.129
sex:diagnosis:feature:region          2.274  4   0.686

ANOVA-Type Statistic (ATS):
                            Test statistic   df1   df2     p-value
sex                                   9.973 1.000 657.416   0.002
diagnosis                            13.124 1.343 657.416   0.000
sex:diagnosis                         1.904 1.343 657.416   0.164
feature                               0.086 1.000     Inf   0.769
sex:feature                           2.167 1.000     Inf   0.141
diagnosis:feature                     1.437 1.562     Inf   0.238
sex:diagnosis:feature                 1.031 1.562     Inf   0.342
region                                0.018 1.611     Inf   0.965
sex:region                            0.371 1.611     Inf   0.644
diagnosis:region                      1.091 2.046     Inf   0.337
sex:diagnosis:region                  0.376 2.046     Inf   0.691
feature:region                        0.126 1.421     Inf   0.810
sex:feature:region                    0.077 1.421     Inf   0.864
diagnosis:feature:region              0.829 1.624     Inf   0.415
sex:diagnosis:feature:region          0.611 1.624     Inf   0.510

p-values resampling:
                                WildBS (WTS) WildBS (ATS)
sex                                 0.000        0.000
diagnosis                           0.000        0.000
sex:diagnosis                       0.119        0.124
feature                             0.798        0.798
sex:feature                         0.152        0.152
diagnosis:feature                   0.067        0.249
sex:diagnosis:feature               0.445        0.362
region                              0.967        0.980
sex:region                          0.691        0.728
diagnosis:region                    0.182        0.338
sex:diagnosis:region                0.863        0.814
feature:region                      0.814        0.926
sex:feature:region                  0.881        0.951
diagnosis:feature:region            0.098        0.519
sex:diagnosis:feature:region        0.764        0.683
\end{CodeOutput}

We find significant effects at level $\alpha=0.01$ of the whole-plot factors sex and diagnosis, while none of the sub-plot factors or interactions become significant.

\subsubsection{Plotting}
The \code{RM()} function is equipped with a plotting option, displaying the calculated means along with $(1-\alpha)$ confidence intervals based on $t$-quantiles. The plot function takes an \code{RM} object as argument. In addition, the factor of interest may be specified. If this argument is omitted in a two- or higher-way layout, the user is asked to specify the factor for plotting. Furthermore, additional graphical parameters can be used to customize the plots. The optional argument \code{legendpos} specifies the position of the legend in higher-way layouts, whereas \code{gap} (default 0.1) is the distance introduced between error bars in a higher-way layout.
\begin{CodeInput}
R> plot(EEG_model, factor = "sex", main = 
+       "Effect of sex on EEG values")
R> plot(EEG_model, factor = "sex:diagnosis", legendpos = "topleft", 
+       col = c(4, 2),  ylim = c(-1.8, 0.8))
R> plot(EEG_model, factor = "sex:diagnosis:feature", 
+       legendpos = "bottomright", gap = 0.05)
\end{CodeInput}

\begin{figure}[h]
	\centering
	\includegraphics[width =0.75\textwidth]{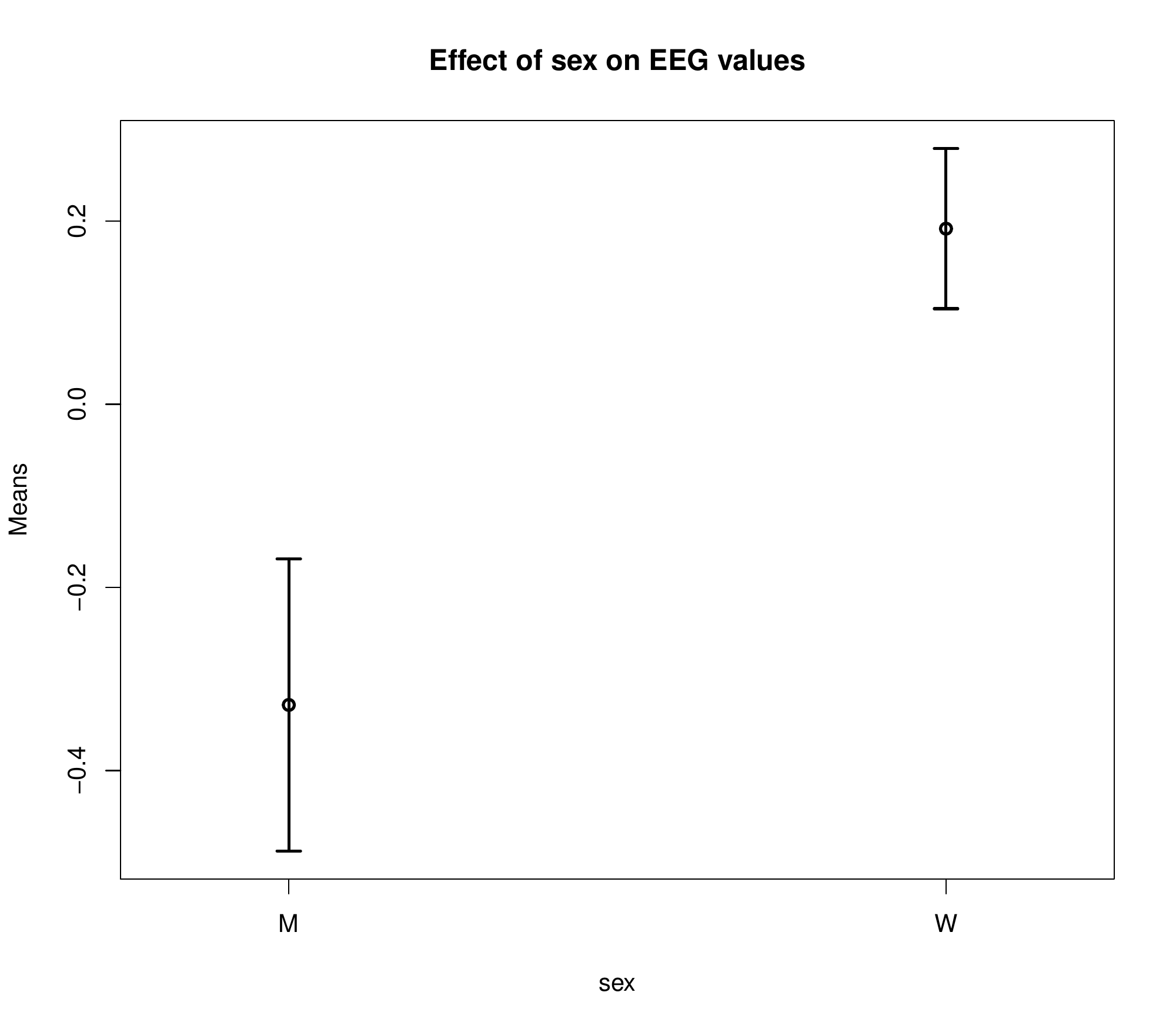}
	\caption{Plot for factor "sex" in the RM model of the EEG data example.}
	\label{Fig:plotting}
\end{figure}

\begin{figure}
	\centering
	\includegraphics[width =0.75\textwidth]{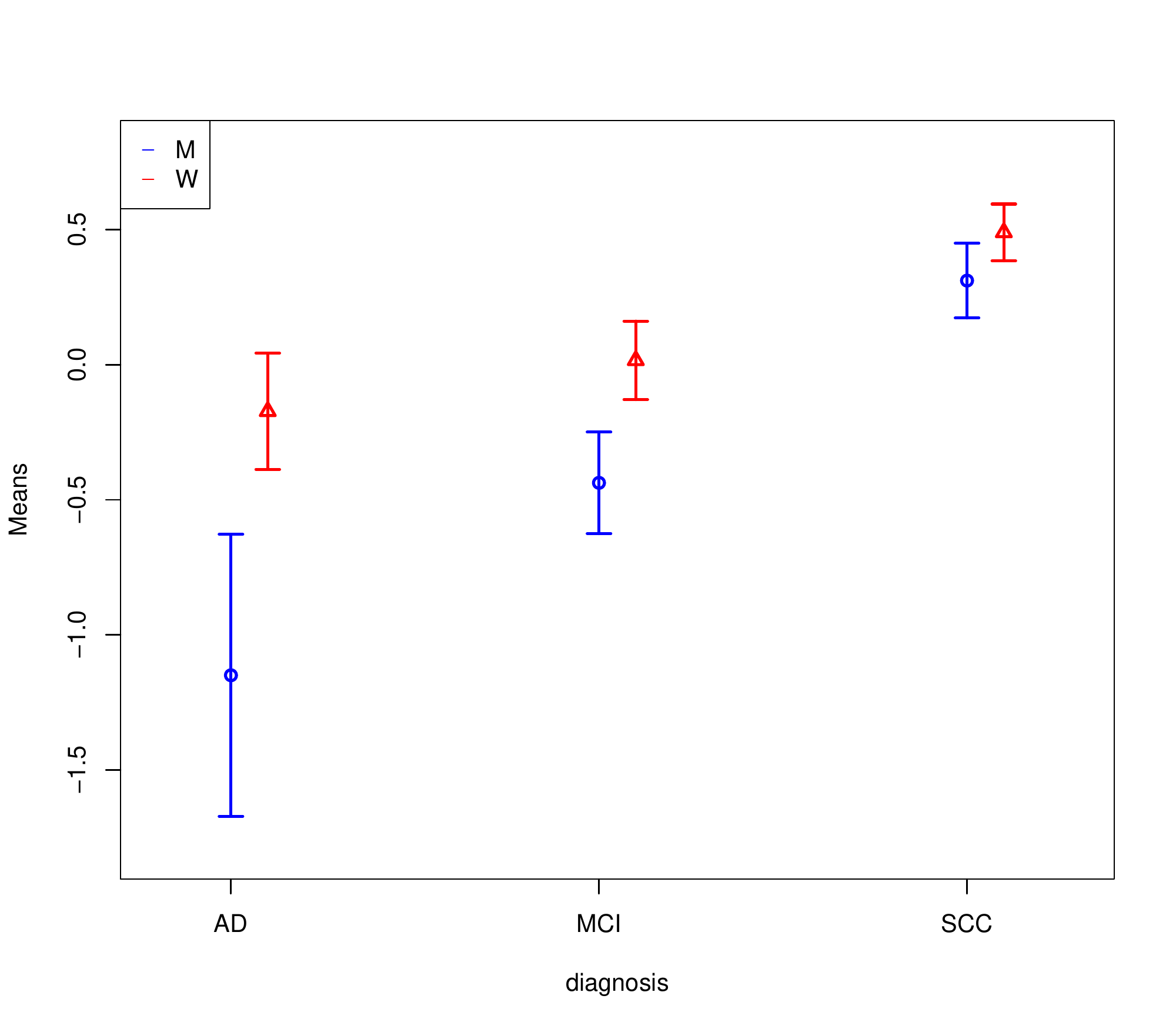}
	\includegraphics[width =0.75\textwidth]{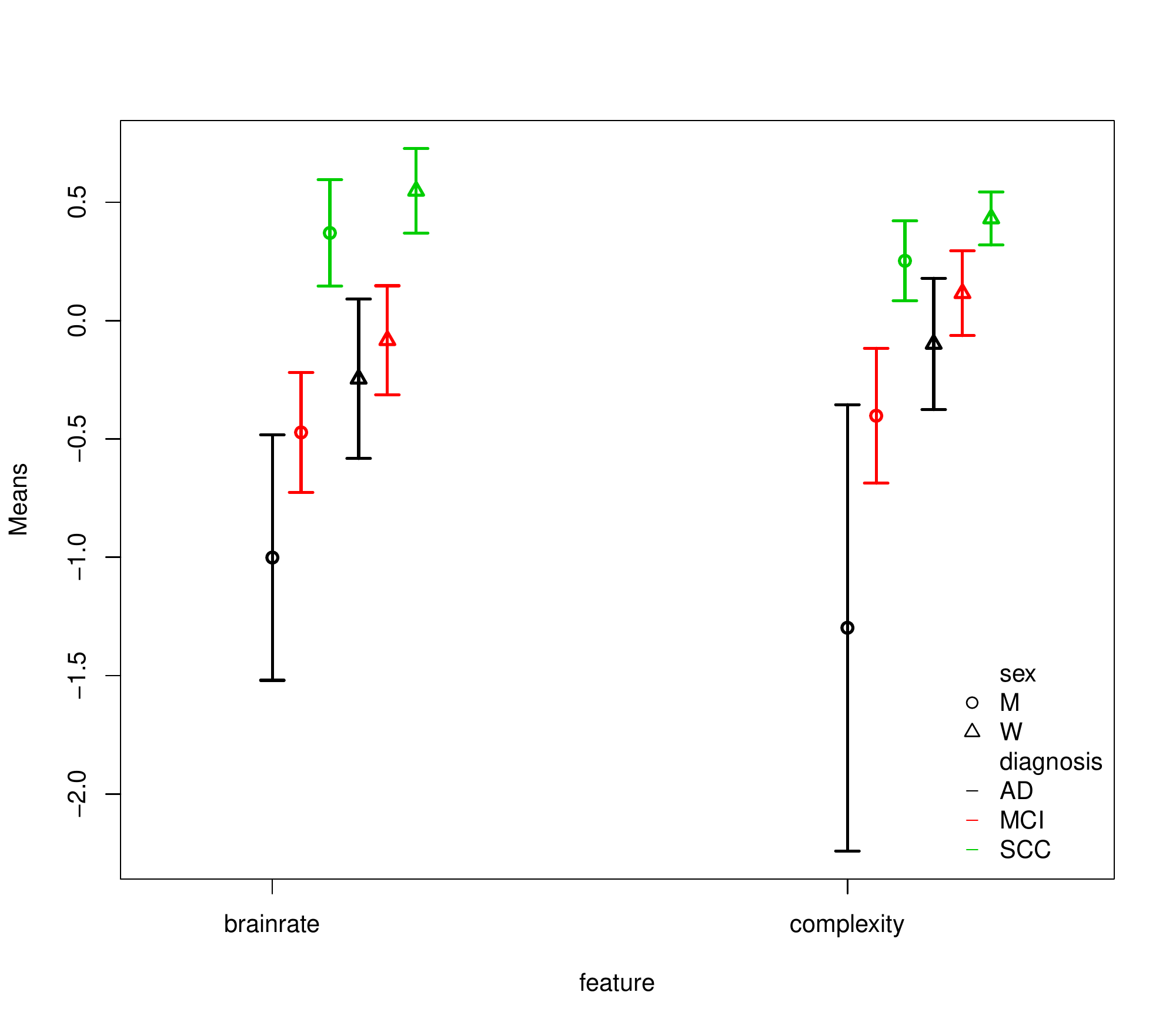}
		\caption{Plot for the interaction between "sex" and "diagnosis" (upper panel) as well as additionally taking "feature" into account (lower panel) in the RM model of the EEG data example.}
		\label{Fig:interactions}
\end{figure}

The resulting plots are displayed in Figure \ref{Fig:plotting} and Figure \ref{Fig:interactions}, respectively.


\subsection{MANOVA Design}

For the analysis of multivariate data, the functions \code{MANOVA} and \code{MANOVA.wide} are implemented. The difference between the two functions is that the response must be stored in long and wide format for using \code{MANOVA} or \code{MANOVA.wide}, respectively. The structure of both functions is very similar. They both calculate the WTS for multivariate data in a design with crossed or nested factors. Additionally, the modified ANOVA-type statistic (MATS) is calculated which has the additional advantage of being applicable to designs involving singular covariance matrices and is invariant under scale transformations of the data \citep{friedrichMATS}. The resampling methods provided are a parametric bootstrap approach and a wild bootstrap using Rademacher weights. Note that only balanced nested designs (i.e., the same number of factor levels $b$ for each level of the factor $A$) with up to three factors are implemented. Designs involving both crossed and nested factors are not implemented.
Note that in nested designs, the levels of the nested factor usually have the same labels for all levels of the main factor, i.e., for each level $i=1,\ldots,a$ of the main factor $A$ the nested factor levels are labeled as $j=1,\ldots,b_i$. If the levels of the nested factor are named uniquely, this has to be specified by setting the parameter \code{nested.levels.unique} to \code{TRUE}.

\begin{CodeInput}
R> MANOVA(formula, data, subject, iter = 10000, alpha = 0.05,
         resampling = "paramBS", CPU, seed,
         nested.levels.unique = FALSE, dec = 3)
R> MANOVA.wide(formula, data, iter = 10000, alpha = 0.05,
              resampling = "paramBS", CPU, seed,
              nested.levels.unique = FALSE, dec = 3)
\end{CodeInput}

The only difference between \code{MANOVA} and \code{MANOVA.wide} in the function call except from the different shape of the formula (see examples below) is the \code{subject} variable, which needs to be specified for \code{MANOVA} only.

\subsubsection{Data Example \code{MANOVA}: Two crossed factors}
We again consider the data set \code{EEG} from the \pkg{MANOVA.RM} package, but now we ignore the sub-plot factor structure. Therefore, we are now in a multivariate setting with 6 measurements per patient and three crossed factors sex, age and diagnosis. Due to the small number of subjects in some groups (e.g., only 2 male patients aged $<$ 70 were diagnosed with AD) we restrict our analyses to two factors at a time. The analysis of this example is shown below.

\begin{CodeInput}
R> data(EEG)
R> EEG_MANOVA <- MANOVA(resp ~ sex * diagnosis, data = EEG,
+	               subject = "id", resampling = "paramBS", 
+	               iter = 1000,  alpha = 0.01, CPU = 1, 
+                       seed = 987)
R> summary(EEG_MANOVA)
\end{CodeInput}

\begin{CodeOutput}
Call: 
resp ~ sex * diagnosis

Descriptive:
  sex diagnosis n  Mean 1 Mean 2  Mean 3 Mean 4 Mean 5 Mean 6
1   M        AD 12 -0.987 -1.007 -1.010 -1.320 -1.086 -1.488
3   M       MCI 27 -0.506 -0.464 -0.447 -0.490 -0.459 -0.257
5   M       SCC 20  0.409  0.243  0.459  0.314  0.095  0.349
2   W        AD 24 -0.285 -0.159 -0.294 -0.194  0.026 -0.128
4   W       MCI 30 -0.069 -0.074 -0.106  0.121  0.131  0.094
6   W       SCC 47  0.559  0.548  0.537  0.506  0.403  0.384

Wald-Type Statistic (WTS):
              Test statistic df  p-value
sex                   12.604  6   0.050
diagnosis             55.158 12   0.000
sex:diagnosis          9.790 12   0.634

modified ANOVA-Type Statistic (MATS):
                Test statistic
sex                   45.263
diagnosis            194.165
sex:diagnosis         18.401

p-values resampling:
                  paramBS (WTS)   paramBS (MATS)
sex                   0.124          0.003
diagnosis             0.000          0.000
sex:diagnosis         0.748          0.223
\end{CodeOutput}

The output consists of several parts: First, some descriptive statistics of the data set are displayed, namely the sample size and mean for each factor level combination and each dimension (dimensions occur in the same order as in the original data set). In this example, Mean 1 to Mean 3 correspond to the brainrate (temporal, frontal, central) while Mean 4--6 correspond to complexity. Second, the results based on the WTS are displayed. For each factor, the test statistic, degree of freedom and $p$-value is given. For the MATS, only the value of the test statistic is given, since here inference is only based on resampling. The resampling-based $p$-values are finally displayed for both test statistics.

To demonstrate the use of the \code{MANOVA.wide()} function, we consider the same data set in wide format, which is also included in the package. In the formula argument, the user now needs to specify the variables of interest bound together via \code{cbind}. A subject variable is no longer necessary, as every row of the data set belongs to one patient in wide format data. The output is almost identically to the one obtained from \code{MANOVA} with the difference that the mean values are now labeled according to the variable names supplied in the \code{formula} argument.

\begin{CodeInput}
R> data("EEGwide")
R> EEG_wide <- MANOVA.wide(cbind(brainrate_temporal, brainrate_frontal,
+           brainrate_central, complexity_temporal, complexity_frontal, 
+           complexity_central) ~ sex * diagnosis, data = EEGwide,
+           resampling = "paramBS",  iter = 1000,  alpha = 0.01,
+           CPU = 1, seed = 987)
R> summary(EEG_wide)
\end{CodeInput}

\begin{CodeOutput}
Call: 
cbind(brainrate_temporal, brainrate_frontal, brainrate_central, 
complexity_temporal, complexity_frontal, complexity_central) ~ 
sex * diagnosis

Descriptive:
  sex diagnosis  n brainrate_temporal  brainrate_frontal  brainrate_central  
1   M        AD 12             -0.987             -1.007             -1.010               
2   W        AD 27             -0.506             -0.464             -0.447          
3   M       MCI 20              0.409              0.243              0.459             
4   W       MCI 24             -0.285             -0.159             -0.294             
5   M       SCC 30             -0.069             -0.074             -0.106              
6   W       SCC 47              0.559              0.548              0.537               
complexity_temporal  complexity_frontal  complexity_central
1      -1.320            -1.086              -1.488
2      -0.490            -0.459              -0.257
3       0.314             0.095               0.349
4      -0.194             0.026              -0.128
5       0.121             0.131               0.094
6       0.506             0.403               0.384

Wald-Type Statistic (WTS):
             Test statistic df p-value
sex                   12.604  6   0.050
diagnosis             55.158 12   0.000
sex:diagnosis          9.790 12   0.634

modified ANOVA-Type Statistic (MATS):
                 Test statistic
sex                   45.263
diagnosis            194.165
sex:diagnosis         18.401

p-values resampling:
                 paramBS (WTS)            paramBS (MATS)
sex                   0.124                     0.003
diagnosis             0.000                     0.000
sex:diagnosis         0.748                     0.223
\end{CodeOutput}

In this example, MATS detects a significant effect of sex, a finding that is not shared by the $p$-value based on the parametric bootstrap WTS.

\subsubsection{Confidence Regions}
The \code{MANOVA} functions are equipped with a function for calculating and plotting of confidence regions. Details on the methods can be found in \citet{friedrichMATS}.
Confidence regions can be calculated using the \code{conf.reg} function. Note that confidence regions can only be plotted in designs with 2 dimensions. 

\begin{CodeInput}
R> conf.reg(object, nullhypo)
\end{CodeInput}

\code{Object} must be an object of class \code{MANOVA}, i.e., created using either \code{MANOVA} or \code{MANOVA.wide}, whereas \code{nullhypo} specifies the desired null hypothesis, i.e., the contrast of interest in designs involving more than one factor.
As an example, we consider the data set water from the \pkg{HSAUR} package \citep{HSAUR}. The data set contains measurements of mortality and drinking water hardness for 61 cities in England and Wales. Suppose we want to analyse whether these measurements differ between northern and southern towns. Since the data set is in wide format, we need to use the \code{MANOVA.wide} function.

\begin{CodeInput}
R> library(HSAUR)
R> data(water)
R> test <- MANOVA.wide(cbind(mortality, hardness) ~ location, 
+          data = water, iter = 1000, resampling = "paramBS",
+          CPU = 1, seed = 123)
R> summary(test)
R> cr <- conf.reg(test)
R> cr
R> plot(cr)
\end{CodeInput}

\begin{CodeOutput}
Call: 
cbind(mortality, hardness) ~ location

Descriptive:
      location  n mortality  hardness
North    North 35  1633.600    30.400
South    South 26  1376.808    69.769



Wald-Type Statistic (WTS):
Test statistic    df        p-value 
51.584          2.000        0.000 

modified ANOVA-Type Statistic (MATS):
      [,1]
[1,] 69.882

p-values resampling:
paramBS (WTS) paramBS (MATS)  
     0              0
\end{CodeOutput}

We find significant differences in mortality and water hardness between northern and southern towns.

\begin{figure}[h]
	\centering
	\includegraphics[width = 0.7\textwidth]{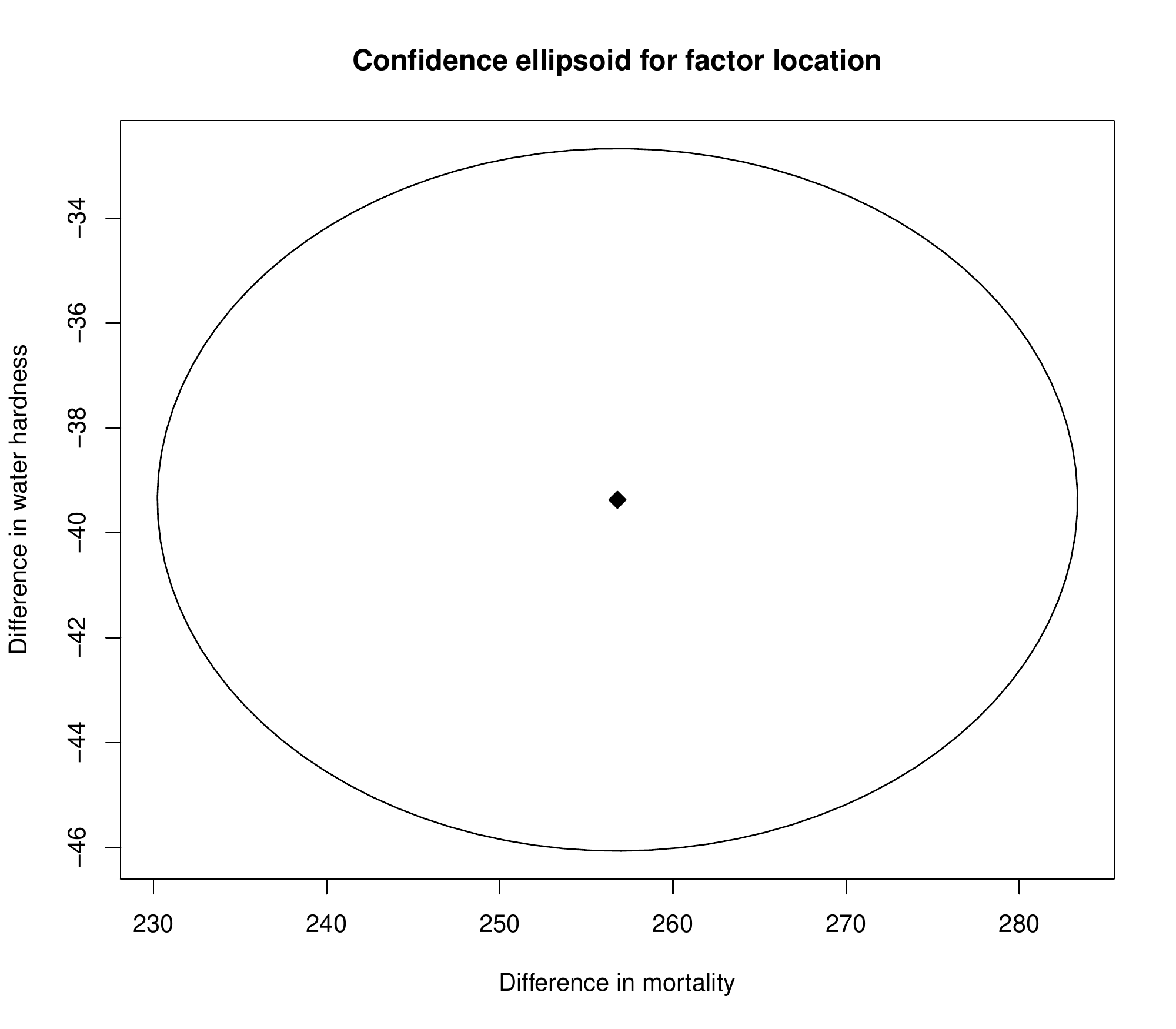}
	\caption{Plot of the confidence region for factor location in the water example from package \pkg{HSAUR}.}
	\label{plot:confreg}
\end{figure}

The confidence region is returned as an ellipsoid specified by its center as well as its axes, which extend \code{Scale} units into the direction of the respective eigenvector. For two-dimensional outcomes as in this example, the confidence ellipsoid can also be plotted, see Figure \ref{plot:confreg}.

\begin{CodeOutput}
Center: 
       [,1]
[1,] 256.792
[2,] -39.369

Scale:
[1] 10.852716  2.736354

Eigenvectors:
      [,1] [,2]
[1,]   -1    0
[2,]    0   -1
\end{CodeOutput}

\subsubsection{Nested design}
To create a data example for a nested design, we use the \code{curdies} data set from the \pkg{GFD} package and extend it by introducing an artificial second outcome variable. In this data set, the levels of the nested factor (site) are named uniquely, i.e., levels 1-3 of factor site belong to "WINTER", whereas levels 4-6 belong to "SUMMER". Therefore, \code{nested.levels.unique} must be set to \code{TRUE}. The code for the analysis using both wide and long format is presented below.

\begin{CodeInput}
R> library(GFD)
R> data(curdies)
R> set.seed(123)
R> curdies$dug2 <- curdies$dugesia + rnorm(36)


	
R> # first possibility: MANOVA.wide
R> fit1 <- MANOVA.wide(cbind(dugesia, dug2) ~ season + season:site,
+          data = curdies, iter = 100, nested.levels.unique = TRUE,
+          seed = 123, CPU = 1)
	
R> # second possibility: MANOVA (long format)
R> dug <- c(curdies$dugesia, curdies$dug2)
R> season <- rep(curdies$season, 2)
R> site <- rep(curdies$site, 2)
R> curd <- data.frame(dug, season, site, subject = rep(1:36, 2))
R> fit2 <- MANOVA(dug ~ season + season:site, data = curd, 
+          subject = "subject", nested.levels.unique = TRUE, 
+          seed = 123, iter = 100, CPU = 1)
	
R> # comparison of results
R> summary(fit1)
R> summary(fit2)
\end{CodeInput}

\begin{CodeOutput}
Call: 
cbind(dugesia, dug2) ~ season + season:site



Descriptive:
  season       site n  dugesia  dug2
1 SUMMER          4 6   0.419 -0.050
2 SUMMER          5 6   0.229  0.028
3 SUMMER          6 6   0.194  0.763
4 WINTER          1 6   2.049  2.497
5 WINTER          2 6   4.182  4.123
6 WINTER          3 6   0.678  0.724

Wald-Type Statistic (WTS):
            Test statistic df p-value
season               6.999  2   0.030
season:site         16.621  8   0.034

modified ANOVA-Type Statistic (MATS):
               Test statistic
season              12.296
season:site         15.064

p-values resampling:
                paramBS (WTS) paramBS (MATS)
season               0.04           0.04
season:site          0.28           0.18

Call: 
dug ~ season + season:site

Descriptive:
  season       site n  Mean 1 Mean 2
1 SUMMER          4 6  0.419 -0.050
2 SUMMER          5 6  0.229  0.028
3 SUMMER          6 6  0.194  0.763
4 WINTER          1 6  2.049  2.497
5 WINTER          2 6  4.182  4.123
6 WINTER          3 6  0.678  0.724
	
Wald-Type Statistic (WTS):
             Test statistic df p-value
season               6.999  2   0.030
season:site         16.621  8   0.034

modified ANOVA-Type Statistic (MATS):
              Test statistic
season              12.296
season:site         15.064

p-values resampling:
               paramBS (WTS) paramBS (MATS)
season               0.04           0.04
season:site          0.28           0.18
\end{CodeOutput}

\subsection{Graphical user interface} \label{sec:GUI}
\begin{figure}[h]
	\centering
		\includegraphics[width=0.975\textwidth]{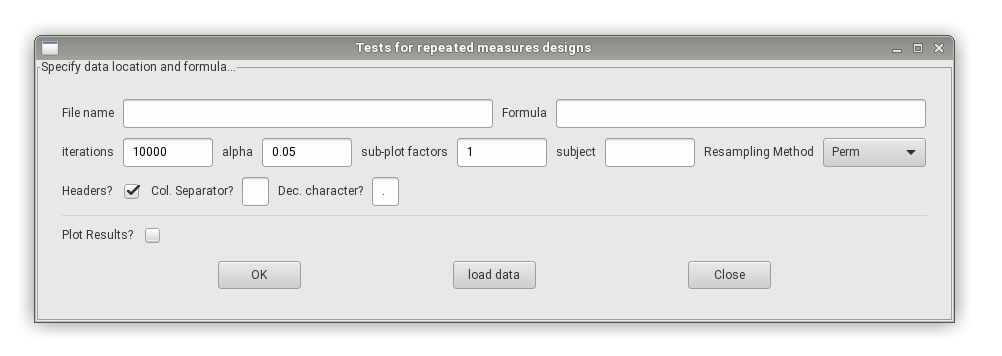}
		\includegraphics[width=\textwidth]{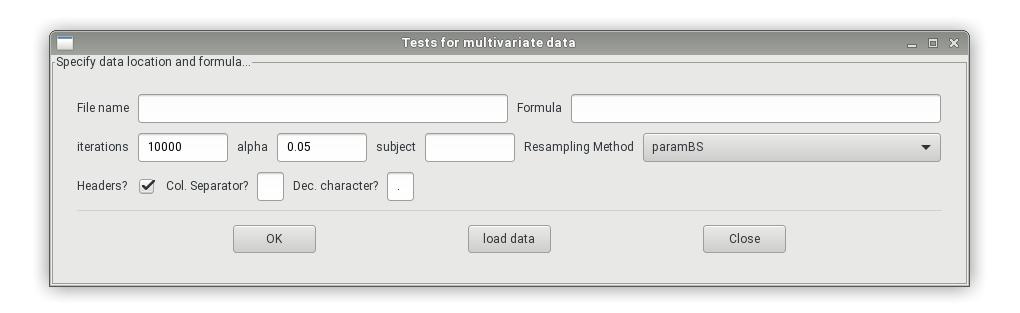}
	\caption{The GUI for tests in repeated measures designs (upper panel) and multivariate data (lower panel): The user can specify the data location and the formula as well as the resampling approach.}
	\label{fig:GUI}
\end{figure}

\begin{figure}[H]
	\centering
		\includegraphics[width=0.37\textwidth]{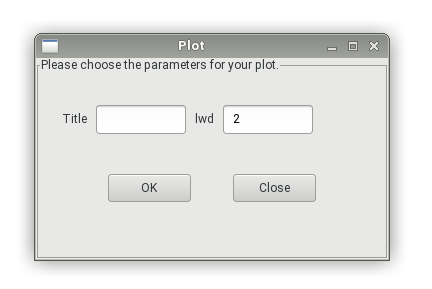} 
		\includegraphics[width=0.56\textwidth]{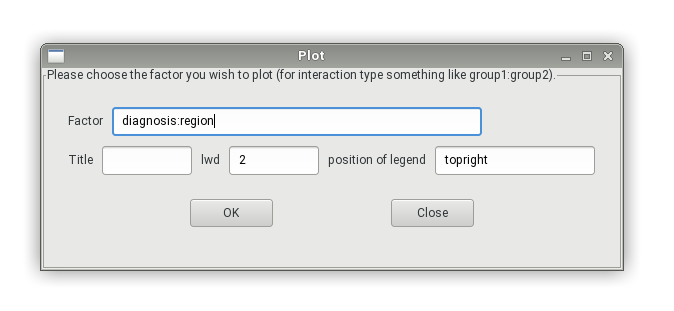}
	\caption{Graphical user interfaces for plotting: The left GUI is for the one-way layout (no choice of factors possible), the right one is for a two-way layout with an example for plotting interactions.}
	\label{fig:GUIplot}
\end{figure}

The GUI is started in \proglang{R} with the command \code{GUI.RM()}, \code{GUI.MANOVA()} and \code{GUI.MANOVAwide()} for repeated measures designs and multivariate data in long or wide format, respectively. {\color{black} Note that the GUI depends on \pkg{RGtk2} and will only work if \pkg{RGtk2} is installed.} The user can specify the data location (either directly or via the "load data" button) and the formula as well as the number of iterations, the significance level $\alpha$, the number of sub-plot factors (for repeated measures designs) and the name of the subject variable, see Figure \ref{fig:GUI}.
Furthermore, the user has the choice between the three resampling approaches "Perm" (only for RM designs), "paramBS" and "WildBS" denoting the permutation procedure, the parametric bootstrap and the wild bootstrap, respectively.
Additionally, one can specify whether or not headers are included in the data file, and which separator and character symbols are used for decimals in the data file. The GUI for repeated measures also provides a plotting option, which generates a new window for specifying the factors to be plotted (in higher-way layouts) along with a few plotting parameters, see Figure~\ref{fig:GUIplot}. 

\begin{CodeInput}
	R> library("MANOVA.RM")
	R> GUI.RM()
\end{CodeInput}

\section{Discussion and Outlook}\label{Conclusion}

We have explicitly described the usage of the R package \pkg{MANOVA.RM} for analyzing various multivariate MANOVA and RM designs. Moreover, the corresponding models and inference procedures that have been derived and theoretically analyzed in previous papers are explained as well. In particular, three different test statistics of Wald-, ANOVA- and modified ANOVA-type are implemented together with appropriate critical values derived from asymptotic considerations, approximations or resampling. Here, the latter is recommended in case of small to moderate sample sizes. All methods can be applied {\it without} assuming usual presumptions such as multivariate normality or specific covariance structures. Moreover, all procedures are particularly constructed to tackle covariance matrix heterogeneity across groups or even covariance singularity (in case of the MATS). In this way \pkg{MANOVA.RM} provides a flexible tool box for inferring hypotheses about (i) main and interaction effects in general factorial MANOVA and (ii) whole- and sub-plot effects in RM designs with possibly complex factorial structures on both, whole- and sub-plots.

In addition, we have placed a graphical user interface (GUI) at the users disposal to allow for a simple and intuitive use. It is planned to update the package on a regular basis; respecting the development of new procedures for general RM and MANOVA designs.
For example, our working group is currently investigating the implementation of covariates in the above model in theoretical research and the resulting procedure may be incorporated in the future. Other topics include the possible implementation of subsequent multiple comparisons, e.g. by the closure principle. 

\section*{Acknowledgments}
The work of Sarah Friedrich and Markus Pauly was supported by the German Research Foundation project DFG-PA 2409/3-1.

\bibliography{Literatur}{}
\bibliographystyle{apalike}

\end{document}